# Emotion-Aware Transformer Encoder for Empathetic Dialogue Generation

Raman Goel, Seba Susan, Sachin Vashisht and Armaan Dhanda
*Department of Information Technology, Delhi Technological University*,
New Delhi, India-110042
seba_406@yahoo.in

*Abstract*—Modern day conversational agents are trained to emulate the manner in which humans communicate. To emotionally bond with the user, these virtual agents need to be aware of the affective state of the user. Transformers are the recent state of the art in sequence-to-sequence learning that involves training an encoder-decoder model with word embeddings from utterance-response pairs. We propose an emotion-aware transformer encoder for capturing the emotional quotient in the user utterance in order to generate human-like empathetic responses. The contributions of our paper are as follows: 1) An emotion detector module trained on the input utterances determines the affective state of the user in the initial phase 2) A novel transformer encoder is proposed that adds and normalizes the word embedding with emotion embedding thereby integrating the semantic and affective aspects of the input utterance 3) The encoder and decoder stacks belong to the Transformer-XL architecture which is the recent state of the art in language modeling. Experimentation on the benchmark Facebook AI empathetic dialogue dataset confirms the efficacy of our model from the higher BLEU-4 scores achieved for the generated responses as compared to existing methods. Emotionally intelligent virtual agents are now a reality and inclusion of affect as a modality in all human-machine interfaces is foreseen in the immediate future.

## I. INTRODUCTION

Virtual assistants such as Cortana, Alexa and Siri, are now an integral part of modern life. Their services range from facilitating simple financial transactions to conducting a friendly conversation with the user. The latter is a more challenging task since the virtual agent needs to emulate the manner in which a human listener empathizes with the emotions of the speaker and generates an appropriate response that syncs with the emotional state of the speaker. To empathize with a human, the machine needs to be emotionally intelligent and needs to be trained on the emotion metadata available in modern dialogue corpora.

Virtual conversational agents generally have at their core a sequence-to-sequence (Seq2Seq) model trained on input-output sentence pairs [1]. In order to generate an appropriate response to a user utterance, the contextual information in the input sentence needs to be captured [2]. Language and sequence modelling tasks require long-term dependencies to be captured effectively from sequential data in order to make sense of the context and identify the meaning conveyed by the sentence. Practically, it is not possible to learn long-term dependencies from an infinitely long time window. Instead, the effort is to learn a long dependency from a time-step that is as large as possible. Recurrent Neural Networks (RNN) such as Long Short-Term Memory (LSTM) [3] and Gated Recurrent Unit (GRU) [4] capture long-term dependencies in sequential data. The standard vanilla RNN suffers from problems like vanishing gradient and exploding gradient. Despite contributing to improvements in machine translation tasks, their performance dropped when the length of the input sentence is increased [4]. Attention mechanism [5] was incorporated into the encoder-decoder architecture for capturing context in the encoder output. Alternatives to purely recurrent encoder-decoder architectures [4, 5] are purely convolutional [6] and convolutional-recurrent architectures [7]. A convolutional encoder was preferred in [7] to better capture the context information in the input sentence in order to generate semantically meaningful responses.

The advent of transformers brought about a revolution in the field of sequence modeling that was mainly dominated by RNNs. The transformer architecture was proposed by Vaswani *et al.* in 2017 [8] for machine translation tasks, that allowed for significant parallel computations and is based entirely on self-attention which was further improved with the "multi-head" attention mechanism. Self-attention allows for better word encoding as each word is processed by the model by looking at the other words in the input sentence which helps in getting better clues/hints about the context in which the current word is being used. The transformer architecture also uses positional encoding by making changes to the original input embedding such that the model learns the position of each word and meaningful distances between different words in the sentence. However, the transformer works with a fixed-length context. This restricts its ability to model dependencies beyond a segment. There is also the problem of context fragmentation which arises due to fixed-length segments not respecting the sentence boundaries leading to a loss of context of the original sentence due to semantic boundaries of the sentence being ignored.

The Transformer-XL model [9] was introduced by Dai *et al.* in 2019 as an improvement over the vanilla transformer with much faster evaluation speeds and much longer dependency learning capability. Transformer-XL (XL for extra-long) overcomes the shortcomings of the vanilla transformer of Al-Rfou *et al.* (2019) [10] by incorporating recurrence mechanism and relative positional encoding. In this paper, we propose a Transformer-XL architecture for empathetic response generation by adding and normalizing the word embedding from the input utterance with the emotion embedding. An emotion detection module forms the first stage in our learning framework. The result is an emotionally intelligent conversational agent that can understand the emotions of the user and generates empathetic responses accordingly. The rest of the paper is organized as follows:

section II presents related work, the proposed model is described in section III, the experimental setup and the results are discussed in sections IV and V respectively, and the paper is concluded in section VI.

## II. RELATED WORK

Most of the contemporary work on conversational agents focus on improving language understanding. Recent works advocate the incorporation of contextual knowledge through attention mechanisms [2, 5]. However, such models are unaware of the emotions of the user, and the generated responses may sometimes reveal the lack of personality or character of the virtual agent. Incorporation of affect as a modality in human-machine interactions is now regarded as a means for gaining human trust. Understanding the affective state of the user is useful for generating emotive responses that improves the overall quality of the human machine interaction and increases customer satisfaction [30]. In two separate works, Lee *et al.* (2005) and Devillers and Vidrascu (2006) presented some initial work on emotion recognition from conversations using audio and lexicon-based methods [11, 12]. Emotion classification using raw audio features is already well explored in literature [13]. Inclusion of text cues improved the classification scores in [14].

Some of the earliest attempts to create emotionally intelligent conversational agents was based on selecting an emotional response using manually crafted rules [31] that required expertise and was not scalable to larger datasets. The introduction of the encoder-decoder model for sequence-to-sequence learning [4] changed the drift of research with most of the researchers adopting this methodology for mapping the input utterance to a suitable output response which is predicted on a word-by-word basis using neural networks. LSTM which is a recurrent neural network captures the temporal information in the words in a sentence effectively [15], hence most of the sequence-to-sequence models have LSTM as the encoder and the decoder. Hu *et al.*, (2018) proposed an encoder-decoder architecture with LSTM to develop a tone-aware chatbot meant for social media customer care [16]. Most of the corpora for dialogue systems, till recently, did not include meta-data such as topic of discussion, speaker-name, personality or emotions. Now, there are available datasets, that are emotionally annotated, like the Facebook AI Empathetic Dialogue (ED) dataset introduced by Rashkin *et al.* in 2019 [17]. There are also datasets based on textual and acoustic features [18, 19] from which models have been developed for sentiment classification and emotion recognition. Zhou *et al.* (2020) used a Seq2Seq GRU-RNN model to develop an empathetic social chatbot which takes into consideration the Emotional Quotient (EQ) and Intelligence Quotient (IQ) for generating responses with proper emotions [20]. Li *et al.* (2019) developed a conversational agent that generated meaningful emotional replies using Reinforcement Learning and emotional editing constraints [21].

An empathetic response-based chatbot called CARO was introduced in [22] that was capable of engaging in empathetic conversations and providing medical advice for people suffering with major depression; it generated responses by appending emotions as a prefix for each utterance. A classification stage forms the first stage of this chatbot that decided whether the user utterance belonged to the category of medical question answering or an empathetic dialogue. The model was trained using both the datasets in this case. Ma *et al.* (2020), has presented a systematic review of chatbots that generate empathetic responses to user utterances; these were termed as Emotionally Aware Chatbots (EAC) [23]. LSTM with attention mechanism has proved to generate more empathetic responses than the LSTM sans attention, as proved by researchers in [24] who experimented on the Facebook AI ED dataset. Likewise, transformer models have proved to outperform LSTM with attention for empathetic dialogue generation, as proved in [17]. The details of our model, which is based on one of the latest advancements in transformer architectures, are described in the next section.

## III. PROPOSED MODEL

Motivated by the work on transformer based empathetic dialogue generation in [17], we propose a novel technique for rendering the virtual agent emotionally intelligent. Assuming that the dialogue dataset has emotion annotations, a separate classifier is trained on the emotion meta-data, apart from the usual training on utterance-response pairs. Fig. 1 shows the basic blocks of our model. To make the dialogue generator emotionally intelligent, we incorporate an emotional bias in the input embedding. The transformer encoder that encodes the input utterance is made emotion aware by incorporating the emotion information in the form of an embedding vector. Unlike [17] that prepends the emotion information along with the utterance embedding as a separate piece of information, we add and normalize each word embedding in the input utterance with the embedding vector of the predicted emotion; the result is provided as input to the transformer encoder. The Keras word embedding is used, that yields better results than the conventional bag-of-words representation used for text classification [25].

Our learning model, therefore, comprises of two independent models that are trained separately on the same dataset; one is the Emotion Classifier and the other is the Generative Chatbot, as shown in Fig. 1. The Emotion Classifier uses the utterance to detect the emotion in the utterance, while the Generative Chatbot provides the utterance and detected emotion as the inputs to the transformer encoder in order to generate the empathetic response. The result is an emotionally intelligent conversational agent that can detect and respond to the emotional state of the user. The details of Emotion Classifier and the Generative Chatbot are given next.

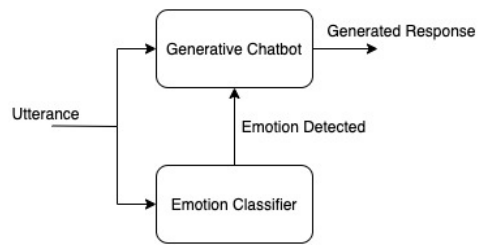

Fig. 1. Basic outline of the proposed model

### A. Emotion Classifier

The first stage of our learning framework comprises of an emotion detection module. Fig. 2(a) shows the distribution of 32 emotions in the ED dataset. We have divided the total number of emotions into eight groups by grouping similar emotions together, as shown in Fig. 2 (b). We used the same coarse grouping of emotions as in [22].

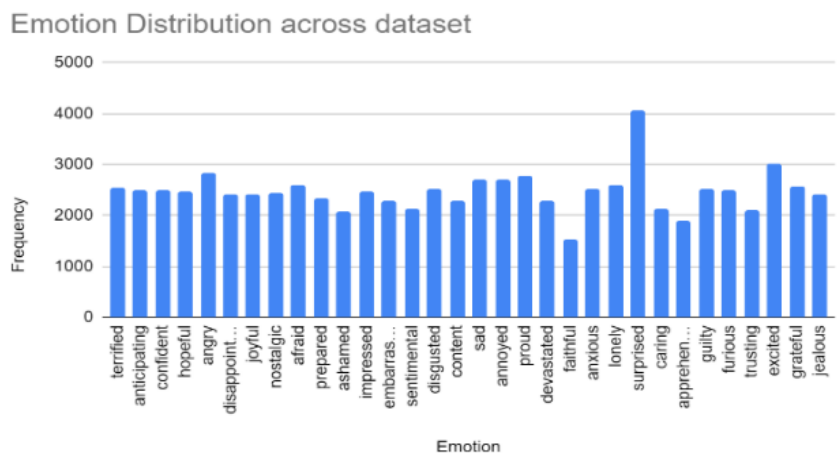

(a)

| Emotion Category | Grouped Emotions |
|---|---|
| excited | excited, surprised, joyful |
| afraid | afraid, terrified, anxious, apprehensive |
| disgusted | disgusted, embarrassed, guilty, ashamed |
| annoyed | angry, annoyed, jealous, furious |
| grateful | faithful, trusting, grateful, caring, hopeful |
| disappointed | sad, disappointed, devastated, lonely, nostalgic, sentimental |
| impressed | proud, impressed, content |
| prepared | anticipating, prepared, confident |

(b)

Fig. 2. (a) Frequency Distribution of emotions in the Facebook AI ED dataset (b) Grouping of emotions in the ED dataset into eight categories (Harilal *et al.* 2020) [22]

The Emotion Classifier that we have used for predicting the context of the given utterance consists of an LSTM unit of 300 cells and a 100-unit dense layer. The softmax layer consists of eight units corresponding to the eight emotion groups. The highest probability in the output layer determines the emotion. The pipeline for our emotion prediction module is shown in Fig. 3.

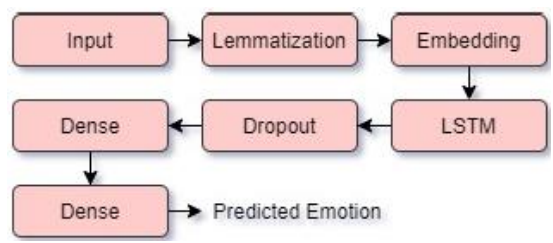

Fig. 3. Emotion classifier

The model uses the dropout regularization technique, in order to prevent model overfitting. The classifier uses the Categorical Cross-Entropy Loss function along with Adam Optimizer [26] for model training.

### B. Generative Chatbot

For the Generative Chatbot, we make use of the Transformer-XL (XL for extra-long) architecture proposed by Dai *et al.* (2019) [9] to model long-term dependencies, which is a recently introduced transformer architecture for language modelling. Transformer-XL has a similar base as that of the vanilla Transformer in [10]. The ability of the Transformer-XL model to handle long-term dependencies in a much better way than the vanilla transformer makes the Transformer-XL much more preferable. In the Transformer-XL architecture, the output of the hidden layer of the last segment is also passed with the present segment's hidden layer output which helps to capture the long term dependencies in a better manner. Also, to capture the long-term dependencies more effectively, positional encoding is introduced in each part of the attention module.

The Generative Chatbot model consists of two parts i.e. encoder and decoder. The encoder receives as its input the word embedding of the utterance augmented by the emotion embedding, as shown in Fig. 4. The embedding vector we get after adding and normalizing the two components has the information about both the words and the emotional context in the input utterance, which will be further processed to generate suitable empathetic responses. The normalization procedure comprises of subtracting the mean and dividing by the standard deviation. An emotional bias is, therefore, introduced in the embedding stage, that makes the transformer encoder emotion aware.

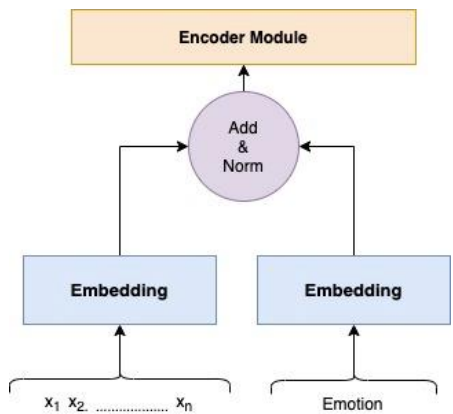

Fig. 4. Augmentation of word embedding with emotion embedding for the transformer encoder

Adding and normalizing embedded feature vectors was found to improve accuracy in several Seq2Seq models due to the induced bias that makes the model sensitive to context [7]. The transformer encoder consists of several layers, each consisting of multi-head attention layer and feed-forward network. The decoder receives the hidden representation from the encoder and decodes it to probabilistically generate the target response, one word at a time. The output sequence obtained from the transformer decoder is the generated response. The responses generated by our model are evaluated for their correctness, as per procedure explained in sections IV and V.

## IV. EXPERIMENTAL SETUP

### A. Dataset

The Facebook AI Empathetic Dialogue (ED) dataset [17] is used for the experimentation. The training dataset comprises of a total of 79189 entries with 8 columns consisting of non-null values of prompt and utterance, associating a prescribed emotion in the context value. A total of 32 unique emotions have been classified in the dialogues that are grouped into 8 emotion categories, as previously shown in Fig. 2. Also, it is quite evident that the 32 different emotions need not specifically highlight unique emotions in real life. For example, '*afraid*' and '*terrified*' can relate to the same emotion type in all practical scenarios. We use the grouping followed by [22] for the ED dataset. A total of 5242 entries are present in the test dataset. It was observed that the relative count of emotions in the train and test dataset is almost the same.

### B. Training setup

The standard text pre-processing steps were performed such as tokenization, converting to lower case. This was followed by lemmatization to standardize the vocabulary. The software implementation was done in Python. We have made our code available online[1]. We use an 8-head Transformer-XL encoder with inner-size dimensionality of 256 and the hidden-layer size is 100. The training process consisted of 50 epochs with a batch size of 64 and a learning rate of 0.001, and the dropout regularization parameter was set to 0.1. Sparse Categorical Cross Entropy loss function was used along with Adam optimizer with the hyperparameters $\beta_1 = 0.9$, $\beta_2 = 0.98$ and $\varepsilon = 10^{-9}$. The loss value was reduced to 0.6054 at the end of 50 epochs.

## V. RESULTS AND DISCUSSIONS

The Emotion Classifier in the first stage of our learning framework achieved a test accuracy of 95%. The predicted emotion was mapped into a 256-dimensional embedded vector, that was added and normalized with the word embedding of the same size. Keras word embedding was used. We used the BLEU-4 score [27] metric to evaluate the quality of the generated responses. This is a value between 0 to 1 that matches the machine generated response with the human response (ground-truth). A value of 1 means a perfect match with the human annotated response. The higher the score, the closer is the machine generated response to the ground-truth. The BLEU-4 scores of all the ground truths available for an input utterance were averaged.

We have considered seven models for comparison, the first is the one proposed by Facebook AI [17] based on the transformer architecture. This model prepends the emotion with the user utterance. The second is the baseline Seq2Seq encoder-decoder model based on LSTM with Attention Mechanism [24]. The third is another LSTM-based conversational agent called CARO [22]. The fourth is the vanilla transformer model proposed by Vaswani *et al.* (2017) [8]. The fifth method EmoDS is that of Song *et al.* (2019) [28], that plugs emotional words at certain time-steps in between words to make the system emotion aware. The sixth method is EmpTransfo [29], a multi-head transformer model that predicts the next emotion and utterance given the current emotion and utterance. The BLEU scores for the responses generated by all models are summarized in Table I.

TABLE I. BLEU SCORES FOR DIFFERENT MODELS

| Models | BLEU Score |
|---|---|
| Facebook AI Model | 0.08 |
| Seq2Seq with Attention Model | 0.137 |
| CARO Model | 0.179 |
| Transformer Model | 0.173 |
| EmoDS Seq2Seq model | 0.173 |
| EmpTransfo | 0.1592 |
| Proposed Model | 0.225 |

As observed, the BLEU score for the proposed model is the highest (0.225), thus, this model generates the most empathetic responses among all methods. However, it was observed that substitution of Keras word embedding with GloVe word embedding reduced the BLEU score of our model from 0.225 to 0.213. A comparison of the responses generated by our model vs. the standard transformer model is shown for a set of utterance-response pairs of the ED dataset in Table II. From the empathetic responses generated by our model, it is evident that it is more emotionally intelligent than the vanilla transformer model. For example, the model is aware of the happy occasion being that of a job, in the first instance, and that the user is going to be a parent, in the second case. The vanilla transformer, on the other hand, generates a generic complimentary response, that has no bearing on the particular situation of the user. The limitation of our model is highlighted in the third case in Table II, where the symbolic notation of math test is not comprehended by both the transformer models and a generic response is generated in both cases that conveys the hope that the problem will be sorted out and that it can happen with anybody.

---
[1] https://github.com/r9729104234/Emotion-Aware-Transformer-Encoder-for-Empathetic-Dialogue-Generation

There are also cases where multiple emotions can be associated with a user utterance. A notable instance is the fourth case in Table II, where the emotion associated with the user statement on skydiving is decoded by our model as "terrified" and a suitable response is generated accordingly. Though the predicted emotion did not match with the dataset annotation ("excited"), the generated response yet was found to sync with some of the ground truth responses that expressed fear and apprehension in regard of skydiving. This leads to the observation that more than one affective state may be involved in human-machine dialogue systems. Identification of compatible human emotions can lead to the development of more emotionally-sensitive human-machine interfaces. Exploration of dynamically evolving emotions in human-machine conversations forms the future scope of our work.

TABLE II. RESPONSES GENERATED BY THE VANILLA TRANSFORMER MODEL VS. THE PROPOSED MODEL FOR SAMPLES FROM THE ED DATASET

| Utterance (user input) | Ground Truth (Human annotations) (Gray highlighted regions introduced to distinguish between ground truths) | Original emotion | Detected Emotion (Coarse classification) | Vanilla Transformer response | Proposed Model response |
|---|---|---|---|---|---|
| I was very happy to get my new job | oh wow. we're practically neighbors. <br><br> I started a new job about a month ago and it's going really well. <br><br> I work in I.T. for the provincial government in Canada | excited | joyful (excited) | That is huge. | Congratulations. What do you do? |
| My first baby is on the way. I'm going to make sure they have a great life. | Oh wow! Congratulations! <br><br> I do not yet but hope to one day. I think you will be a great parent! <br><br> Thank you. We have been trying for a year. | grateful | hopeful (grateful) | Congratulations! That is wonderful. | Oh wow! I think you will be a great parent |
| I was hoping that I would do better on a math test. | My math teacher told the class I needed to do better on my next test. <br><br> Well you should try to study more and see what happens <br><br> Well maybe you should talk about getting a tutor or something | grateful | hopeful (grateful) | It surely can be sorted. | It can happen to anyone. |
| I'm going skydiving this weekend. | Oh my! That sounds exciting for sure! I couldn't do it, I don't think. <br><br> I would say that you're an adventurous person. Me, I'm more laid back and mellow. Kudos to you and have fun! <br><br> That's i'm so scared but it will be worth it. You only have one life so don't hold back. <br><br> I'm going skydiving this weekend. I'm afraid of heights and freaking out a little. | excited | terrified (afraid) | Have a great weekend. It's fun. | Have you ever done anything else so frightening? |

## VI. CONCLUSIONS

In this paper, we have presented an emotionally intelligent conversational agent that is capable of conducting empathetic conversations with users. The proposed model integrates both the semantic and affective aspects of the input utterance. The two embeddings (semantic, affective) are added and normalized and fed as input to the transformer encoder. For the same set of utterances, the BLEU score improved from 0.173 to 0.225 as we moved from the standard transformer to the advanced Transformer-XL architecture, which is also reflected in the difference in the quality of the responses generated by the two models. The proposed conversational agent successfully emulates a human agent that can perceive user emotion and respond empathetically to the user input. Emotionally intelligent virtual agents are now a reality and the inclusion of affect as a modality in all human-machine interfaces is foreseen in the immediate future.